\documentclass[aps,prc,showpacs,twocolumn,groupedaddress]{revtex4}

\usepackage{graphics}

\begin{document}

%\preprint{}

\title{Electron screening in the liquid-gas mixed phases of nuclear matter}


\author{Gentaro Watanabe}
%\thanks{}
\affiliation{
Department of Physics, University of Tokyo, 7-3-1 Hongo, Bunkyo,
Tokyo 113-0033, Japan \\ 
and Division of Computational Science, 
The Institute of Physical and Chemical Research (RIKEN), 2-1 Hirosawa,
Wako, Saitama 351-0198, Japan}
\author{Kei Iida}
%\thanks{}
\affiliation{
The Institute of Physical and Chemical Research (RIKEN), 2-1 Hirosawa,
Wako, Saitama 351-0198, Japan}


\date{\today}

\begin{abstract}

     Screening effects of electrons on inhomogeneous nuclear matter, which 
includes spherical, slablike, and rodlike nuclei as well as spherical and 
rodlike nuclear bubbles, are investigated in view of possible application to 
cold neutron star matter and supernova matter at subnuclear densities.  Using 
a compressible liquid-drop model incorporating uncertainties in the surface 
tension, we find that the energy change due to the screening effects broadens 
the density region in which bubbles and nonspherical nuclei appear in the 
phase diagram delineating the energetically favorable shape of inhomogeneous 
nuclear matter.  This conclusion is considered to be general since it stems 
from a model-independent feature that the electron screening acts to decrease
the density at which spherical nuclei become unstable against fission and to 
increase the density at which uniform matter becomes unstable against proton 
clustering.

\end{abstract}

%\keywords{}

\maketitle

\section{\label{intro} Introduction}

     At subnuclear densities and low temperatures where nuclei are so closely 
packed in a gas consisting mainly of dripped neutrons as to melt into uniform 
matter, nuclear matter is expected to possess spatially periodic structures 
composed of rodlike and slablike nuclei and rodlike and spherical bubbles 
\cite{review}; these nuclei and bubbles are often referred to as nuclear 
``pasta.''  Seminal works by Ravenhall $et$ $al$. \cite{rpw} and Hashimoto $et$
$al$. \cite{hashimoto} based on liquid-drop models show that a subtle balance 
between nuclear surface and Coulomb energies determines the 
energetically most favorable ``pasta'' shape.
Such ``pasta''-like structures may be encountered in the outer part of a 
stellar remnant of collapse-driven supernova explosion.  In this part, the 
inhomogeneous nuclear matter would be neutralized and roughly 
$\beta$-equilibrated by a relativistic degenerate gas of electrons and, 
during and just after collapse, of electron neutrinos.  [Hereafter, such 
material will be denoted as supernova matter or neutron star matter, 
according to whether the degenerate neutrino gas coexists or not.]
The charge screening 
action of the electron gas on clumps of protons reduces the Coulomb energy 
induced by the protons while increasing the kinetic energy of the electron 
gas.  These energy corrections, which have yet to be examined in detail, may 
affect the density region in which the ``pasta'' phases are preferred over the
low density phase with roughly spherical nuclei and the high density phase of
uniform matter.

     In this paper we ask the question of how the electron screening changes 
the energies of zero-temperature supernova matter and neutron star matter at 
subnuclear densities.  For the purpose of answering this question, it is 
instructive to consider the 
two opposite limits: perfect screening and no screening.  In the case of 
perfect screening where the electrostatic energy vanishes and the kinetic 
energy of the electron gas is maximal, a coexisting state of a nucleon liquid 
of density comparable to the normal nuclear density $n_0\simeq0.16$ fm$^{-3}$ 
and a low density neutron gas can occur at a unique pressure as a phase 
separated state.  On the other hand, in the absence of charge screening, i.e.,
in the homogeneous limit of the electron gas, the electrostatic energy is 
maximal while the electron kinetic energy is minimal.  In the liquid region
in $\beta$ equilibrium,
the decrease in the electron Fermi energy as compared to the perfect screening 
case plays a role in increasing the proton fraction and thus lowering the bulk
energy of the nucleon liquid.  Due to such decrease in the electron and nucleon
bulk energy, a liquid-gas coexisting state can occur for a finite range of 
pressures as a mixed state composed of alternating liquid-gas regions.  At 
sufficiently low densities, this mixed state is expected to manifest itself as
a state in which the liquid regions correspond to roughly spherical nuclei and
the gas regions are almost vacant.  At densities just below $n_0$, as 
suggested by earlier studies \cite{review}, such a mixed state may appear as
the ``pasta'' phases mentioned above, depending on uncertain quantities such 
as the neutron and proton chemical potentials of a neutron gas and the surface 
tension in the presence of dripped neutrons.  A real situation -- {\em partial}
screening -- lies between those two limiting cases, whereas 
it remains to be clarified how such screening affects the density region of 
the ``pasta'' phases through modifications of the bulk and electrostatic 
energies.

     The key quantity of the electron screening is the ratio of the scale of 
a nucleus or bubble to the Thomas-Fermi screening length 
$\lambda_{\rm TF}^{(e)}$ as given by
\begin{equation}
  \lambda_{\rm TF}^{(e)}= \kappa_{e}^{-1}
     = \left[
       4\pi e^{2}\left( \frac{\partial n_{e}^{(0)}}{\partial \mu_{e}^{(0)}}
       \right)_{n_{e}^{(0)}}\right]^{-1/2}\ ,
\end{equation}
where $n_{e}^{(0)}$ and $\mu_{e}^{(0)}$ are the averaged electron number 
density and chemical potential, respectively.
If the ratio is much smaller than unity, the electron density can be assumed 
to be everywhere constant to a good approximation.  The typical values of 
$\lambda_{\rm TF}^{(e)}$ at subnuclear densities for neutron star matter
(upper) and supernova matter (lower) 
can be estimated in the massless limit as
\begin{eqnarray}
\lambda_{\rm TF}^{(e)} &=&
\sqrt{\frac{\pi}{4\alpha}}\ (k_e^{(0)})^{-1}\nonumber\\
&\simeq& \left\{
  \begin{array}{rl}
    20~ \mbox{fm} & (n \simeq 0.5n_{0}, x \simeq 0.1),\\
    15~ \mbox{fm} & (n \simeq 0.5n_{0}, x \simeq 0.3),
  \end{array}
\right.
\end{eqnarray}
where $\alpha$ is the fine structure constant, $k_e^{(0)} = (3\pi^{2} 
n_e^{(0)})^{1/3}$ is the electron Fermi wave number, and $n$ and $x$ are the 
nucleon density and proton fraction in the nucleon liquid.
These values of $\lambda_{\rm TF}^{(e)}$ are larger than the typical values 
of half thickness of a nucleus or bubble, $\sim 5$ fm, and even half the 
internuclear spacing, $\sim 10$--15 fm (see, e.g., Refs.\ \cite{lorenz,
oyamatsu,williams,lassaut,gentaro1,gentaro2}).
Consequently, the 
standard approximation in which no screening is included seems fairly valid.
As we shall show within the linearized Thomas-Fermi approximation 
\cite{dyson}, the screening correction to the density region in which the 
phases with bubbles and nonspherical nuclei appear is indeed small for the 
typical values of the surface tension, while 
the screening makes
such a density region 
larger rather than narrower.  This feature will be concluded to be general
by investigating how the electron screening affects the conditions for 
instabilities against fission of spherical nuclei and proton clustering in 
uniform matter.

     Quantum molecular dynamics simulations that have recently been performed 
by one of the authors successfully reproduce the ``pasta'' phases expected to
occur as the ground state, without imposing any assumptions on nuclear shapes 
\cite{qmd}.  In this framework, the electron screening has not yet been taken 
into consideration, whereas proton screening, i.e., Coulombic polarization of
the nuclear interior, is automatically incorporated.  Before examining the 
influence of the electron screening on the simulations, it is useful to focus 
on the qualitative nature of the electron screening effects, which will be 
clarified in the present work.  We remark in passing that the proton screening
makes the liquid portion deviate only slightly from uniformity, although the 
proton screening length is comparable to the scale of the liquid region.

     To investigate the electron screening effects on the phase diagram for
the ground-state neutron star matter and supernova matter at subnuclear 
densities, we extend a compressible liquid-drop model of inhomogeneous nuclear
matter immersed in the lepton gas as constructed in Refs.\ \cite{gentaro1} and
\cite{gentaro2} by incorporating the electron density deviation from the 
non-screening case in the Thomas-Fermi approximation up to linear order in the
electrostatic potential \cite{dyson}.  This model, a generalization of the one
developed by Baym, Bethe, and Pethick \cite{bbp} (hereafter BBP), allows us to
obtain the phase diagram in a way dependent on uncertain values of the surface
tension.  It is essential to take into account such uncertainties; the surface
tension is related to the equilibrium size of a nucleus or bubble, which in 
turn determines the efficiency of the screening.

     This paper is composed as follows.  In Sec.\ \ref{model}, we construct a 
compressible liquid-drop model for nuclei and bubbles, and write down 
equilibrium conditions for zero-temperature neutron star matter and supernova 
matter.  Expressions for the electrostatic energy including the electron 
screening effects are derived in Sec.\ \ref{tf}.  The influence of the 
electron screening on the phase diagram is discussed in Sec.\ \ref{result}.

\section{\label{model} Model of Matter at Subnuclear Densities}

     In this section, we provide the free energy and equilibrium conditions 
for zero-temperature neutron star matter and supernova matter at densities of 
order 0.1--1$n_0$.  The zero-temperature approximation is considered to well 
describe the energy of matter in the deepest region of neutron star crusts 
since the temperature of the crusts is typically $\alt 10^{9}$K, which is much
smaller than typical nuclear and electronic excitation energies.  
For densities considered here, on the other hand, matter under stellar 
collapse is well characterized by the lepton fraction $Y_{\rm L}$, 
\begin{equation}
  Y_{\rm L}=\frac{n_{e}^{(0)}+n_{\nu}}{n_{\rm b}}\ ,\label{yl}
\end{equation}
where $n_{\nu}$ is the number density of a uniform gas of electron 
neutrinos, and $n_{\rm b}$ is the averaged baryon density.  The lepton fraction
typically amounts to 0.3--0.38 due to the trapping of electron neutrinos 
\cite{freedman,sato}.  Hereafter, we set $Y_{\rm L}=0.3$.  Since the 
degeneracy pressure of the neutrino gas is much larger than the thermal
pressure 
at temperatures
of order 10-50 MeV, to a first approximation, we shall include the 
energy of the degenerate neutrino gas into the energy of zero-temperature 
neutron star matter, and ignore finite temperature effects.

     Following Refs.\ \cite{gentaro1} and \cite{gentaro2}, we consider five 
phases that consist of spherical nuclei, cylindrical nuclei, planar nuclei, 
cylindrical bubbles, and spherical bubbles, respectively.  Each phase is 
assumed to be composed of a Coulomb lattice of a single species of nucleus or 
bubble at a given baryon density $n_{\rm b}$.  We adopt the Wigner-Seitz 
approximation in evaluating the lattice energy with sufficient accuracy 
\cite{oyamatsu}.  In this approximation, a cell in the bcc lattice, including 
a spherical nucleus or bubble of radius $r_{\rm N}$, is replaced by a 
Wigner-Seitz cell defined as a sphere having radius $r_{\rm c}$ and the same 
center.  A cylindrical nucleus or bubble having an infinitely long axis and a 
circular section of radius $r_{\rm N}$ is taken to be contained in a 
cylindrical Wigner-Seitz cell having the same axis and a circular section of 
radius $r_{\rm c}$ in place of a cell in the two-dimensional triangular 
lattice.  For a planar nucleus with thickness $2r_{\rm N}$, a Wigner-Seitz 
cell is identical with a cell in the one-dimensional layered lattice, having 
width $2r_{\rm c}$.  The values of $r_{\rm c}$ for these phases are chosen so 
that each Wigner-Seitz cell may have zero net charge.

     Hereafter we shall concentrate on the charge screening of proton clumps 
by the electron gas.  We shall ignore curvature effects, 
nucleon pairing effects, shell effects in inhomogeneous nuclear matter 
\cite{OY,magierski}, and fluctuation-induced displacements of nuclei and 
bubbles \cite{gentaro1,gentaro2}, which may have significant consequence
to the spatial structure of nuclear matter.

\subsection{Energy of matter}

      A compressible liquid-drop model for nuclei and bubbles \cite{review} is
useful for investigating the nature of the electron screening.  While we 
assume neutrons and protons to be distributed uniformly inside and outside the
nuclei or bubbles and electron neutrinos to be an ideal uniform gas as in 
earlier investigations, we newly include 
the screening-induced deviation, $\delta n_{e}({\bf r})$, of the electron 
number density from the unperturbed constant value $n_{e}^{(0)}$; 
$\delta n_{e}({\bf r})$ will be explicitly calculated within the linear 
Thomas-Fermi approximation in the next section.
Calculating the contributions up to second order in $\delta n_{e}({\bf r})$ 
accurately, we may write the total energy density, $E_{\rm tot}$, averaged 
over a single cell as
\begin{equation}
E_{\rm tot}=\left\{
            \begin{array}{ll}
                w_{\rm N}+w_{\rm L}+(1-u)E_{n}(n_{n})
                +E_{e}[n_{e}({\bf r})]&\\
                \ +E_{\nu}(n_{\nu})
                   & \mbox{(nuclei)}\ , \\ \\
                w_{\rm N}+w_{\rm L}+uE_{n}(n_{n})
                +E_{e}[n_{e}({\bf r})]&\\
                \ +E_{\nu}(n_{\nu})
                   & \mbox{(bubbles)}\ .
            \end{array}\right.
\label{etot}
\end{equation}
Here $w_{\rm N}$ is the energy of the nuclear matter region (the region 
containing protons) in a cell as divided by the cell volume, $w_{\rm L}$ is 
the lattice energy per unit volume, $n_{n}$ is the number density of dripped 
neutrons outside the nuclei or inside the bubbles, $E_{n}$, $E_{e}$,
and $E_{\nu}$ are the energy densities of the neutron matter, of
the electron gas, and of the neutrino gas, respectively,
and $u$ is the volume fraction occupied by the nuclei or bubbles:
\begin{equation}
u = \left(\frac{r_{\rm N}}{r_{c}} \right)^d = \left\{
    \begin{array}{ll}
        \displaystyle{\frac{n_{\rm b}-n_{n}}{n-n_{n}}}
        & \quad \mbox{(nuclei)\ ,}\smallskip\\
        \displaystyle{\frac{n-n_{\rm b}}{n-n_{n}}}
        & \quad \mbox{(bubbles)\ ,}
    \end{array}\right.
\end{equation}
where $d$ is the dimensionality defined as $d=1$ for slabs, $d=2$ for 
cylinders, and $d=3$ for spheres, and $n$ is the nucleon number density inside 
the nuclear matter region.
Note that the contribution of the density deviation $\delta n_e({\bf r})$ is 
included in Eq.\ (\ref{etot}) through the lattice energy $w_{\rm L}$ and the 
electron energy $E_e$.

     The expressions for $E_{n}$, $E_{e}$, and $E_{\nu}$ in
Eq.\ (\ref{etot}) are given by
\begin{eqnarray}
  E_{n}(n_{n})&=&[W(k_{n},0) + m_{n}c^{2}]\ n_{n}\ , \\
  E_{e}(n_{e}^{(0)})&=&\frac{3}{4} \hbar c k_{e}^{(0)} n_{e}^{(0)}\ ,\\
  E_{e}[n_{e}({\bf r})]&=&\frac{1}{V_{\rm c}}\int_{\rm cell}d^{3}{\bf r}
        \left[E_{e}(n_{e}^{(0)})
          +\frac{\partial E_{e}(n_{e}^{(0)})}{\partial n_{e}}\ \delta n_{e}
          \right.\nonumber\\ &&\left.
          +\frac{1}{2}\frac{\partial^{2}E_{e}(n_{e}^{(0)})}{\partial n_{e}^{2}}
          \ \delta n_{e}^{2} + O(\delta n_{e}^{3}) \right]\ ,\\
        &\simeq& E_{e}(n_{e}^{(0)}) \left[ 1+\frac{2}{9 V_{\rm c}}
          \int_{\rm cell}\left(\frac{\delta n_{e}({\bf r})}{n_{e}^{(0)}}
          \right)^{2}d^{3}{\bf r}\right]\ ,\qquad\\
  E_{\nu}(n_{\nu})&=&\frac{3}{4} \hbar c k_{\nu} n_{\nu}\ ,
\end{eqnarray}
where $W(k_{n},0)$ with $k_{n}=(3\pi^{2}n_{n}/2)^{1/3}$ is the energy per 
neutron for uniform neutron matter, $V_{\rm c}$ is the cell volume, and
$k_{\nu}=(6\pi^{2}n_{\nu})^{1/3}$ is the neutrino Fermi wave number.

     For the energy of the nuclear matter region, $w_{\rm N}$, we adopt a 
generalized version of the compressible liquid-drop model developed by BBP 
\cite{bbp}, which gives rise to \cite{gentaro1}
\begin{eqnarray}
&&w_{\rm N}(n,x,n_{n},r_{\rm N},r_{\rm c},d)\nonumber\\
&&\quad = \left\{
\begin{array}{l}
un[(1-x)m_{n}+xm_{p}]c^2+unW(k,x)\\
\quad+w_{\rm surf}(n,x,n_{n},r_{\rm N},u,d)+w_{\rm C}(n,x,r_{\rm N},u,d)
\\ \hspace{55mm} 
 \mbox{(nuclei)\ ,}\\ \\
(1-u)n[(1-x)m_{n}+xm_{p}]c^2\\ 
\quad+(1-u)nW(k,x)+w_{\rm surf}(n,x,n_{n},r_{\rm N},u,d)\\
\quad+w_{\rm C}(n,x,r_{\rm N},u,d)
\\ \hspace{55mm} 
 \mbox{(bubbles)\ ,}
\end{array}\right.
\end{eqnarray}
where $m_{n}\ (m_{p})$ is the neutron (proton) rest mass, $W(k,x)$ is the 
energy per nucleon for uniform nuclear matter of nucleon Fermi wave number 
$k=(3\pi^{2}n/2)^{1/3}$ and proton fraction $x$, as given by BBP [see Eq.\ 
(3.19) in Ref.\ \cite{bbp}], $w_{\rm surf}$ is the nuclear surface energy per 
unit volume, and $w_{\rm C}$ is the self Coulomb energy (per unit volume) of 
protons contained in a cell.

     This expression for $w_{\rm N}$ includes three parameters $C_{1}$, 
$C_{2}$, and $C_{3}$, which are associated with uncertainties in the proton 
chemical potential $\mu_{p}^{(0)}$ in pure neutron matter as contained in 
$W(k,x)$ and those in the nuclear surface tension 
\begin{equation}
 E_{\rm surf}=r_{\rm N}w_{\rm surf}/ud\ .
 \label{esurf}
\end{equation}
The parameter
$C_{1}$ determines the magnitude of $\mu_{p}^{(0)}$ (not including the rest 
mass) as [Eq.\ (4) in Ref.\ \cite{gentaro1}]
\begin{equation}
\mu_{p}^{(0)}=-C_1 n_{n}^{2/3}\ .
\end{equation} 
Hereafter we shall set $C_{1}=400$ MeV fm$^{2}$; this case is consistent with
the $n_n$ dependence of $\mu_{p}^{(0)}$ obtained from various model 
calculations as exhibited in Fig.\ 1 of Ref.\ \cite{gentaro1}.  The other two 
parameters $C_{2}$ and $C_{3}$ are defined as [Eq.\ (6) in Ref.\ 
\cite{gentaro1}]
\begin{equation}
E_{\rm surf}=C_{2}\tanh\left(\frac{C_{3}}{\mu_{n}^{(0)}}\right) 
E_{\rm surf}^{\rm BBP}\ ,
\end{equation}
where $\mu_{n}^{(0)}=\partial E_{n}/\partial n_{n} - m_{n}c^2$ is 
the neutron chemical potential in the neutron gas (not including the rest 
mass), and $E_{\rm surf}^{\rm BBP}$ is the BBP-type surface tension [Eq.\ (7) 
in Ref.\ \cite{gentaro1}].  As shown in Fig.\ 1 of Ref.\ \cite{gentaro1}
and Fig.\ 1 of Ref.\ \cite{gentaro2}, the 
surface tension $E_{\rm surf}$ calculated for $C_{2}=1.0$ and $C_{3}=3.5$ MeV 
agrees well with the Hartree-Fock results obtained by Ravenhall, Bennett, and 
Pethick \cite{rbp} using a Skyrme interaction.  In the present work, we thus
assume $C_{3}=3.5$ MeV, while we give $C_2$ a range of values including unity,
0.1--10. 
To set such various values for the parameter $C_2$ determining the
strength of $E_{\rm surf}$ allows us to investigate the electron screening
effects more clearly since the typical size of the nuclear matter region 
approaches the screening length as $C_{2}$ becomes larger.

\subsection{Equilibrium conditions}

     Zero-temperature neutron star matter with nuclei or bubbles of given 
shape, in its equilibrium, fulfills the conditions for stability of the 
nuclear matter region against change in the size, neutron drip, $\beta$-decay, 
and pressurization (see Section 2.2 in Ref.\ \cite{gentaro1}).  These 
conditions arise from minimization of the energy density $E_{\rm tot}$ with 
respect to five variables $n$, $x$, $n_{n}$, $r_{\rm N}$, and $u$ 
at vanishing $n_{\nu}$ and fixed baryon density $n_{\rm b}$ given by
\begin{equation}
n_{\rm b} = \left\{
\begin{array}{ll}
un+(1-u)n_{n} & \quad \mbox{(nuclei)\ ,}\\
(1-u)n+un_{n} & \quad \mbox{(bubbles)\ ,}
\end{array}\right.
\label{nb}
\end{equation}
as well as under charge neutrality,
\begin{equation}
n_{e}^{(0)} = \left\{
\begin{array}{ll}
xnu & \quad \mbox{(nuclei)\ ,}\\
xn(1-u) & \quad \mbox{(bubbles)\ .}
\end{array}\right.
\label{cn}
\end{equation}
In order to obtain the equilibrium conditions for supernova matter, one has
only to repeat the minimization at fixed $Y_{\rm L}$, Eq.\ (\ref{yl}), and 
$n_{\rm b}$, Eq.\ (\ref{nb}), rather than at vanishing $n_{\nu}$ and fixed 
$n_{\rm b}$.

     The expression for the size equilibrium that can be obtained from
optimization of $E_{\rm tot}$ with respect to $r_{\rm N}$ at fixed 
$n,\ x,\ n_{n}$, and $u$ is
\begin{equation}
  \displaystyle{\left.
    \frac{\partial}{\partial r_{\rm N}}\ (w_{\rm surf} + w_{\rm C+L}
    + E_{e}[n_{e}({\bf r})])
  \right|_{n, x, n_{n}, u} = 0}
\label{sizeeq}
\end{equation}
This expression does {\em not} agree with the non-screening formula [see
Eq.\ (14) in Ref.\ \cite{gentaro1}],
\begin{equation}
 w_{\rm surf}=2 w_{\rm C+L}\ ,
\label{usual}
\end{equation}
since the total electrostatic energy density denoted by $w_{\rm C+L} \equiv 
w_{\rm C} + w_{\rm L}$ is no longer proportional to $r_{\rm N}^2$ for fixed 
$u$ (see Sec.\ \ref{tf}).

     The $\beta$-equilibrium condition can be written as
\begin{equation}
\mu_{e} = \mu_{n}^{\rm (N)} - \mu_{p}^{\rm (N)}
   + \mu_{\nu} + (m_{n}-m_{p})c^{2}
\label{betaeq}
\end{equation}
where $\mu_{p}^{\rm (N)}$ is the proton chemical potential in the nuclear 
matter region given by Eq.\ (20) in Ref.\ \cite{gentaro1},
$\mu_{\nu}=\hbar k_{\nu}c$ is the neutrino chemical potential, which vanishes
for neutron star matter, and 
\begin{widetext}
\begin{equation}
  \mu_{e}= 
 \left\{
\begin{array}{ll}
\hbar c k_{e}^{(0)}+\displaystyle{\frac{2}{9}\frac{1}{nuV_{\rm c}}
    \frac{\partial}{\partial x}\left.\left[ E_{e}(n_{e}^{0})
      \int \left(\frac{\delta n_{e}({\bf r})}{n_{e}^{(0)}}\right)^{2}
        d^{3}{\bf r}\right]\right|_{n,n_{n},r_{\rm N},r_{\rm c}}}&
  \mbox{(nuclei)}\ ,
  \smallskip\\
\hbar c k_{e}^{(0)}+\displaystyle{\frac{2}{9}\frac{1}{n(1-u)V_{\rm c}}
    \frac{\partial}{\partial x}\left.\left[ E_{e}(n_{e}^{0})
      \int \left(\frac{\delta n_{e}({\bf r})}{n_{e}^{(0)}}\right)^{2}
        d^{3}{\bf r}\right]\right|_{n,n_{n},r_{\rm N},r_{\rm c}}}&
  \mbox{(bubbles)}\ .\\
\end{array}
\right.\label{mu_e}
\end{equation}
\end{widetext}
is the electron chemical potential.

     The expressions for the remaining two equilibrium conditions are 
unchanged from the non-screening case \cite{gentaro1,gentaro2}.  The condition
for the drip equilibrium reads 
\begin{equation}
 \mu_{n}^{\rm (N)} = \mu_{n}^{\rm (G)}\ , \label{dripeq}
\end{equation}
where $\mu_{n}^{\rm (N)}$ and $\mu_{n}^{\rm (G)}$ are the neutron chemical 
potentials in the nuclear matter region and in the neutron gas given by 
Eqs.\ (16) and (17) in Ref.\ \cite{gentaro1}, respectively.  The pressure 
equilibrium condition can be expressed as
\begin{equation}
P^{\rm (N)} = P^{\rm (G)}\ , \label{preeq}
\end{equation}
where $P^{\rm (N)}$ and $P^{\rm (G)}$ are the pressures of the nuclear matter 
region and of the neutron gas given by Eqs.\ (22) and (23) in Ref.\ 
\cite{gentaro1}, respectively.

     In order to see the difference from the non-screening case explicitly, it
is instructive to write down the expressions for $\mu_{n}^{\rm (N)}$ and 
$P^{\rm (N)}$ that can be obtained after expressing $x$, $r_{\rm N}$, and 
$r_{\rm c}$ in terms of $n_{\rm b}$, $n$, and $n_n$ through the baryon density
(\ref{nb}), the size equilibrium condition (\ref{sizeeq}), and the 
$\beta$-equilibrium condition (\ref{betaeq}).  The neutron chemical potential
$\mu_{n}^{\rm (N)}$ in the nuclear matter region reads
\begin{widetext}
\begin{equation}
\mu_{n}^{\rm (N)}= 
 \left\{
\begin{array}{ll}
\displaystyle{W(k,x)+\frac{k}{3}\frac{\partial W(k,x)}{\partial k}
+ \left.
  \frac{1}{u} \frac{\partial w_{\rm C+L}}{\partial n}
\right|_{x,r_{\rm N},r_{\rm c}}
+ \left.
\frac{d}{r_{\rm N}}\frac{\partial E_{\rm surf}}{\partial n}
\right|_{x,n_{n}}
+ x[\mu_{e}-\mu_{\nu}-(m_{n}-m_{p})c^2]}
& \mbox{(nuclei)\ ,}\smallskip\\
\displaystyle{W(k,x)+\frac{k}{3} \frac{\partial W(k,x)}{\partial k}
+ \left.
  \frac{1}{1-u} \frac{\partial w_{\rm C+L}}{\partial n}
\right|_{x,r_{\rm N},r_{\rm c}}
+ \left.
\frac{d}{r_{\rm N}}\frac{u}{1-u}\frac{\partial E_{\rm surf}}{\partial n}
\right|_{x,n_{n}}
+ x[\mu_{e}-\mu_{\nu}-(m_{n}-m_{p})c^2]}
& \mbox{(bubbles)\ .}
\end{array}\right.
\label{mun}
\end{equation}
The pressure of the nuclear matter region $P^{\rm (N)}$ reads
\begin{equation}
P^{({\rm N})}= \left\{
\begin{array}{ll} 
\displaystyle{\frac{nk}{3} \frac{\partial W(k,x)}{\partial k}
- \frac{d-1}{r_{\rm N}} E_{\rm surf}
+ \left.\frac{dn}{r_{\rm N}} 
  \frac{\partial E_{\rm surf}}{\partial n}\right|_{x,n_{n}}  
- \left.
  \frac{r_{\rm N}}{d u} \frac{\partial}
  {\partial r_{\rm N}}(w_{\rm C+L}+E_{e}[n_{e}({\bf r})])
\right|_{nu, x, r_{\rm c}}}
& \mbox{(nuclei)\ ,}\smallskip\\
\displaystyle{\frac{nk}{3} \frac{\partial W(k,x)}{\partial k}
+ \frac{d-1}{r_{\rm N}} E_{\rm surf}
+ \left.\frac{dn}{r_{\rm N}} \frac{u}{1-u} 
  \frac{\partial E_{\rm surf}}{\partial n}\right|_{x,n_{n}}
+ \left.
  \frac{r_{\rm N}}{d u} \frac{\partial}
  {\partial r_{\rm N}}(w_{\rm C+L}+E_{e}[n_{e}({\bf r})])
\right|_{n(1-u), x, r_{\rm c}}}
& \mbox{(bubbles)\ .}
\end{array}\right.
\label{pn}
\end{equation}
\end{widetext}
For neutron star matter, these expressions are modified from the non-screening
expressions, Eqs.\ (24) and (26) in Ref.\ \cite{gentaro1}, due to the 
difference in the size equilibrium between Eqs.\ (\ref{sizeeq}) and 
(\ref{usual}).

     Finally, we can calculate the equilibrium values of $n,\ x,\ n_{n},\ 
r_{\rm N}$, and $r_{\rm c}$ and thus the optimal value of $E_{\rm tot}$
for neutron star matter (supernova matter) at given $n_{\rm b}$ and nuclear 
shape (at $Y_{\rm L}=0.3$ and at given $n_{\rm b}$ and nuclear shape) by 
incorporating 
Eqs.\ (\ref{mun}) and (\ref{pn}) into the conditions for drip and pressure 
equilibria, Eqs.\ (\ref{dripeq}) and (\ref{preeq}).  
By comparing the optimal values of 
$E_{\rm tot}$ obtained for the five crystalline phases and uniform matter, we 
can determine the phase giving the smallest energy density at various values 
of $n_{\rm b}$ and $C_{2}$ and thereby draw the phase diagram for the ground 
state neutron star matter and supernova matter,
as will be shown in Sec.\ \ref{result}.

\section{\label{tf} Electrostatic Energy}

     We proceed to derive analytic formulas for the electrostatic energy 
density $w_{\rm C+L}$ in the five crystalline phases by taking into account 
the electron screening within the framework of the linearized Thomas-Fermi 
approximation \cite{dyson}.  In this framework, the energy density functional 
$\varepsilon[n_i({\bf r}),\phi({\bf r})]$ ($i=n,p,e,\nu;$ $\phi$, the 
electrostatic potential), which is related to the average energy density 
$E_{\rm tot}$ as $E_{\rm tot}=V_{\rm c}^{-1}\int_{\rm cell} d^3 {\bf r}
\varepsilon[n_i({\bf r}),\phi({\bf r})]$, is expanded up to second 
order in $\delta n_e({\bf r})$.  Minimization of the expanded energy density 
functional with respect to $\phi({\bf r})$ and $\delta n_e({\bf r})$ leads to 
the Poisson equation,
\begin{eqnarray}
  &&\nabla^{2} \phi({\bf r}) - \kappa_{e}^{2} \phi({\bf r})\nonumber\\
  &&\qquad = - 4 \pi n_{Q}({\bf r})\nonumber\\
  &&\qquad = \left\{
    \begin{array}{ll}
      - 4 \pi e[n x \theta(r_{\rm N}-|{\bf r}|) - n_{e}^{(0)}]
      & \mbox{(nuclei)}\ , \\
      - 4 \pi e[n x \theta(|{\bf r}|-r_{\rm N}) - n_{e}^{(0)}]
      & \mbox{(bubbles)}\ ,
    \end{array}\right.\label{poisson}
\end{eqnarray}
where $n_{Q}$ is the local charge density in the non-screening limit, and the 
relation between $\delta n_e({\bf r})$ and $\phi$, 
\begin{equation}
  \delta n_{e}({\bf r}) = \frac{\kappa_{e}^{2}}{4\pi e^{2}} e \phi({\bf r})\ .
    \label{dev}
\end{equation}
For planar configuration, we take Cartesian coordinates in which the $z$-axis 
is normal to the nuclear surface and the origin is located on the central plane
of the nucleus; for cylindrical configuration, cylindrical coordinates in which
the radial coordinate $\rho$ is normal to the surface of the nucleus or bubble
and the line of $\rho=0$ coincides with the symmetry axis of the nucleus or 
bubble; for spherical configuration, spherical coordinates in which the radial
coordinate $r$ is normal to the surface of the nucleus or bubble and the 
origin is located at its center.  In physical realizations, the electrostatic 
potential and its derivative must be continuous throughout the system.  Thus, 
appropriate boundary conditions are that the derivative of the electrostatic 
potential at the origin and the cell boundary is zero, i.e., $\phi'(0)=0$ and 
$\phi'(r_{\rm c})=0$; and that the electrostatic potential and its derivative 
are continuous at the surface of the nucleus or bubble.

      We turn to the solutions to the Poisson equation (\ref{poisson}) for 
slablike nuclei, cylindrical nuclei and bubbles, and spherical nuclei and 
bubbles.  The results for the electrostatic potential read
\begin{itemize}
\begin{widetext}
\item slab 
\begin{equation}
  \phi(z) = \left\{
    \begin{array}{lr}
      \displaystyle{
      \pm\frac{4\pi nxe}{\kappa_{e}^{2}}
      \left[
        - \frac{\sinh[\kappa_{e}(r_{\rm c}-r_{\rm N})]}
        {\sinh(\kappa_{e}r_{\rm c})}\
        \cosh(\kappa_{e}|z|)
        + (1-u)
      \right]}
    & \quad 0 \le |z| \le r_{\rm N} , \\ \\
      \displaystyle{
      \pm\frac{4\pi nxe}{\kappa_{e}^{2}}
      \left[
        \frac{\sinh(\kappa_{e} r_{\rm N})}{\sinh(\kappa_{e} r_{\rm c})}\
        \cosh[\kappa_{e}(r_{\rm c}-|z|)] - u
      \right]}
    & \quad r_{\rm c} \ge |z| > r_{\rm N} ,
    \end{array}\right.\label{phi1}
\end{equation}
\item cylinder
\begin{equation}
  \phi(\rho) = \left\{
    \begin{array}{lr}
      \displaystyle{
      \pm\frac{4\pi nxe}{\kappa_{e}^{2}}
      \left[
        \frac{K_{1}(\kappa_{e}r_{\rm c}) I_{1}(\kappa_{e}r_{\rm N})
          - I_{1}(\kappa_{e}r_{\rm c}) K_{1}(\kappa_{e}r_{\rm N})}
        {I_{1}(\kappa_{e}r_{\rm c})}\
        \kappa_{e}r_{\rm N} I_{0}(\kappa_{e}\rho)
        + (1-u)
      \right]}
    & \quad 0 \le \rho \le r_{\rm N} , \\ \\
      \displaystyle{
      \pm\frac{4\pi nxe}{\kappa_{e}^{2}}
      \left[
        \kappa_{e}r_{\rm N} I_{1}(\kappa_{e}r_{\rm N})
        \left\{
          \frac{K_{1}(\kappa_{e}r_{\rm c})}{I_{1}(\kappa_{e}r_{\rm c})}
          I_{0}(\kappa_{e}\rho) + K_{0}(\kappa_{e}\rho)
        \right\}
        - u
      \right]}
    & \quad r_{\rm c} \ge \rho > r_{\rm N} ,
    \end{array}\right.\label{phi2}
\end{equation}
\item sphere
\begin{equation}
  \phi(r) = \left\{
    \begin{array}{lr}
      \displaystyle{
      \pm\frac{4\pi nxe}{\kappa_{e}^{2}}  }\smallskip\\ 
      \displaystyle{\quad\times
      \left[
        - \frac{\kappa_{e}(r_{\rm c} - r_{\rm N})
          \cosh[\kappa_{e}(r_{\rm c}-r_{\rm N})]
          + (\kappa_{e}r_{\rm c}\kappa_{e}r_{\rm N} - 1)
          \sinh[\kappa_{e}(r_{\rm c}-r_{\rm N})]}
        {\kappa_{e}r_{\rm c}\cosh(\kappa_{e}r_{\rm c})
          - \sinh(\kappa_{e}r_{\rm c})}\
        \frac{\sinh(\kappa_{e}r)}{\kappa_{e}r}
        + (1-u)
      \right]}
    & \quad 0 \le r \le r_{\rm N} , \\ \\
      \displaystyle{
      \pm\frac{4\pi nxe}{\kappa_{e}^{2}}
      \left[
        \frac{\kappa_{e}r_{\rm N} \cosh(\kappa_{e}r_{\rm N})
          - \sinh(\kappa_{e}r_{\rm N})}
        {\kappa_{e}r_{\rm c}\cosh(\kappa_{e}r_{\rm c})
          - \sinh(\kappa_{e}r_{\rm c})}\
        \frac{\kappa_{e}r_{\rm c}\cosh[\kappa_{e}(r_{\rm c}-r)]
          - \sinh[\kappa_{e}(r_{\rm c}-r)]}{\kappa_{e}r}
        - u
      \right]}
    & \quad r_{\rm c} \ge r > r_{\rm N} ,
    \end{array}\right.\label{phi3}
\end{equation}
\end{widetext}
\end{itemize}
where the upper plus (lower minus) sign corresponds to nuclei (bubbles); 
$I_{n}$ and $K_{n}$ are the $n$-th modified Bessel functions.

      Correspondingly, the electric field $E$ can be obtained from the 
derivative of the potential $\phi$ as
\begin{itemize}
\begin{widetext}
\item slab
\begin{equation}
  E(z) = \left\{
    \begin{array}{lr}
      \displaystyle{
      \pm\frac{4\pi nxe}{\kappa_{e}}\
        \frac{\sinh[\kappa_{e}(r_{\rm c}-r_{\rm N})]}
        {\sinh(\kappa_{e}r_{\rm c})}\
        \sinh(\kappa_{e}|z|)\ \frac{z}{|z|}
      }
    & \quad 0 \le |z| \le r_{\rm N} , \\ \\
      \displaystyle{
      \pm\frac{4\pi nxe}{\kappa_{e}}\
        \frac{\sinh(\kappa_{e} r_{\rm N})}{\sinh(\kappa_{e} r_{\rm c})}\
        \sinh[\kappa_{e}(r_{\rm c}-|z|)]\ \frac{z}{|z|}
      }
    & \quad r_{\rm c} \ge |z| > r_{\rm N} ,
    \end{array}\right.\label{e1}
\end{equation}
\item cylinder
\begin{equation}
  E(\rho) = \left\{
    \begin{array}{lr}
      \displaystyle{
      \pm\frac{4\pi nxe}{\kappa_{e}}\
        \frac{I_{1}(\kappa_{e}r_{\rm c}) K_{1}(\kappa_{e}r_{\rm N})
          - K_{1}(\kappa_{e}r_{\rm c}) I_{1}(\kappa_{e}r_{\rm N})}
        {I_{1}(\kappa_{e}r_{\rm c})}\
        \kappa_{e}r_{\rm N} I_{1}(\kappa_{e}\rho)
      }
    & \quad 0 \le \rho \le r_{\rm N} , \\ \\
      \displaystyle{
      \pm\frac{4\pi nxe}{\kappa_{e}}\
      \kappa_{e}r_{\rm N} I_{1}(\kappa_{e}r_{\rm N})
      \left[
          K_{1}(\kappa_{e}\rho)
          - \frac{K_{1}(\kappa_{e}r_{\rm c})}{I_{1}(\kappa_{e}r_{\rm c})}
          I_{1}(\kappa_{e}\rho)
      \right]}
    & \quad r_{\rm c} \ge \rho > r_{\rm N} ,
    \end{array}\right.\label{e2}
\end{equation}
\item sphere
\begin{equation}
  E(r) = \left\{
    \begin{array}{lr}
      \displaystyle{
      \pm\frac{4\pi nxe}{\kappa_{e}}\
        \frac{\kappa_{e}(r_{\rm c} - r_{\rm N})
          \cosh[\kappa_{e}(r_{\rm c}-r_{\rm N})]
          + (\kappa_{e}r_{\rm c}\kappa_{e}r_{\rm N} - 1)
          \sinh[\kappa_{e}(r_{\rm c}-r_{\rm N})]}
        {\kappa_{e}r_{\rm c}\cosh(\kappa_{e}r_{\rm c})
          - \sinh(\kappa_{e}r_{\rm c})}\  }\smallskip\\ \displaystyle{\qquad\times
        \frac{\kappa_{e}r\cosh(\kappa_{e}r) - \sinh(\kappa_{e}r)}
        {(\kappa_{e}r)^{2}}
      }
    & \quad 0 \le r \le r_{\rm N} , \\ \\
      \displaystyle{
      \pm\frac{4\pi nxe}{\kappa_{e}}\
        \frac{\kappa_{e}r_{\rm N} \cosh(\kappa_{e}r_{\rm N})
          - \sinh(\kappa_{e}r_{\rm N})}
        {\kappa_{e}r_{\rm c}\cosh(\kappa_{e}r_{\rm c})
          - \sinh(\kappa_{e}r_{\rm c})}\  }\smallskip\\ \displaystyle{\qquad\times
        \frac{\kappa_{e}(r_{\rm c}-r)\cosh[\kappa_{e}(r_{\rm c}-r)]
          + (\kappa_{e}r_{\rm c}\kappa_{e}r - 1)
          \sinh[\kappa_{e}(r_{\rm c}-r)]}{(\kappa_{e}r)^{2}}
      }
    & \quad r_{\rm c} \ge r > r_{\rm N} ,
    \end{array}\right.\label{e3}
\end{equation}
\end{widetext}
\end{itemize}
where the upper plus (lower minus) sign corresponds to nuclei (bubbles) as in 
Eqs.\ (\ref{phi1})--(\ref{phi3}).

     We remark that the electron density deviation $\delta n_e$, which can be 
obtained by substituting the solutions for $\phi$ given by Eqs.\
(\ref{phi1})--(\ref{phi3}) into Eq.\ (\ref{dev}) does not violate charge 
neutrality in a cell.  This is because $\int_{\rm cell} \delta n_{e}\ 
d^{3}{\bf r} \propto \int_{\rm cell} \phi\ d^{3}{\bf r} = 0$.

     The total electrostatic energy $W_{\rm C+L}$ in a cell is given by
\begin{equation}
  W_{\rm C+L}
  = \frac{1}{8\pi} \int_{\rm cell}E^{2} d^{3}{\bf r}\ .
  \label{electrostatic}
\end{equation}
The electrostatic energy density $w_{\rm C+L}$ can thus be calculated by 
substituting Eqs.\ (\ref{e1})--(\ref{e3}) into Eq.\ (\ref{electrostatic}) as
\begin{equation}
  w_{\rm C+L} = 2\pi(nxe)^{2} r_{\rm N}^{2} u
     f_{d}^{\rm (screen)}(\kappa_{e}r_{\rm c},\kappa_{e}r_{\rm N})\ ,
\end{equation}
with
\begin{itemize}
\begin{widetext}
\item slab
\begin{eqnarray}
  f_{1}^{\rm (screen)}(s,t) &=&
  \frac{1}{2t^{2}} \left\{
    -\frac{s}{t} \frac{\sinh^{2}t}{\sinh^{2}s}
    +\frac{1}{t} \frac{\sinh t}{\sinh s} \sinh(s-t)
    +2 \frac{\sinh t}{\sinh s} \cosh(s-t) -1 \right\}\ ,\label{f1}
\end{eqnarray}
\item cylinder
\begin{eqnarray}
  f_{2}^{\rm (screen)}(s,t) &=&
  I_{1}^{2}(t) \left[
    \frac{K_{1}(t)}{I_{1}(t)} \left\{
      \frac{I_{1}(t)}{K_{1}(t)} K_{0}(t)K_{2}(t)
      - \frac{K_{1}(t)}{I_{1}(t)} I_{0}(t)I_{2}(t)
      -2\frac{K_{1}(s)}{I_{1}(s)} \left(
        I_{1}^{2}(t) - I_{0}(t)I_{2}(t) \right)
    \right\}
  \right.\nonumber\\
  &&\qquad\left.
    +\left(\frac{s}{t}\right)^{2} K_{1}^{2}(s) \left\{
      2 - \frac{I_{0}(s)I_{2}(s)}{I_{1}^{2}(s)}
      - \frac{K_{0}(s)K_{2}(s)}{K_{1}^{2}(s)}
    \right\}
  \right.\nonumber\\
  &&\qquad\left.
    -\frac{1}{t^{2}} \frac{K_{1}(s)}{I_{1}(s)} \frac{1}{\sqrt{\pi}} \left\{
      G_{24}^{22}\left(s\left|
          \begin{array}{llll}
          1,&\frac{3}{2}& & \\ 1,&2,&0,&0
          \end{array}\right)\right.
      - G_{24}^{22}\left(t\left|
          \begin{array}{llll}
          1,&\frac{3}{2}& & \\ 1,&2,&0,&0
          \end{array}\right)\right.
    \right\}
  \right]\ ,\label{f2}
\end{eqnarray}
\item sphere
\begin{eqnarray}
  f_{3}^{\rm (screen)}(s,t) &=&
  \frac{3}{2 t^{6}}
  \frac{1}{(s \cosh s - \sinh s)^{2}}
  \nonumber\\
  &&\qquad
  \times\biggl[\biggl\{-\left(1+(s-t) (s^{2}t-s-t)\right)
      + \bigl(1+s(s-t)\bigr) \cosh[2(s-t)]
  \nonumber\\
    &&\qquad\qquad\left.\left.
      + \frac{1}{2} \left(s^{2}t-4s+t \right) \sinh[2(s-t)]
    \right\}
    \left(t\cosh t - \sinh t\right)^{2}
    \right.\nonumber\\
    &&\qquad
    + \left\{(s-t) \cosh(s-t) + (st-1) \sinh(s-t)\right\}^{2}
    \left(t^{2}+ t \cosh t\ \sinh t - 2\sinh^{2}t \right)
  \biggr]\ ,\label{f3}
\end{eqnarray}
\end{widetext}
\end{itemize}
where $G_{24}^{22}$ is the Meijer's G-function defined as
\begin{eqnarray}
  G_{24}^{22}\left(z\left|
        \begin{array}{llll}
        1,&\frac{3}{2}& & \\ 1,&2,&0,&0
        \end{array}\right)\right.
    &\equiv& 4\sqrt{\pi} \int I_{1}(z) K_{1}(z) z dz\ .\qquad
\end{eqnarray}

\begin{figure}[htbp]
\begin{center}
\rotatebox{0}{
\resizebox{7.5cm}{!}
{\includegraphics{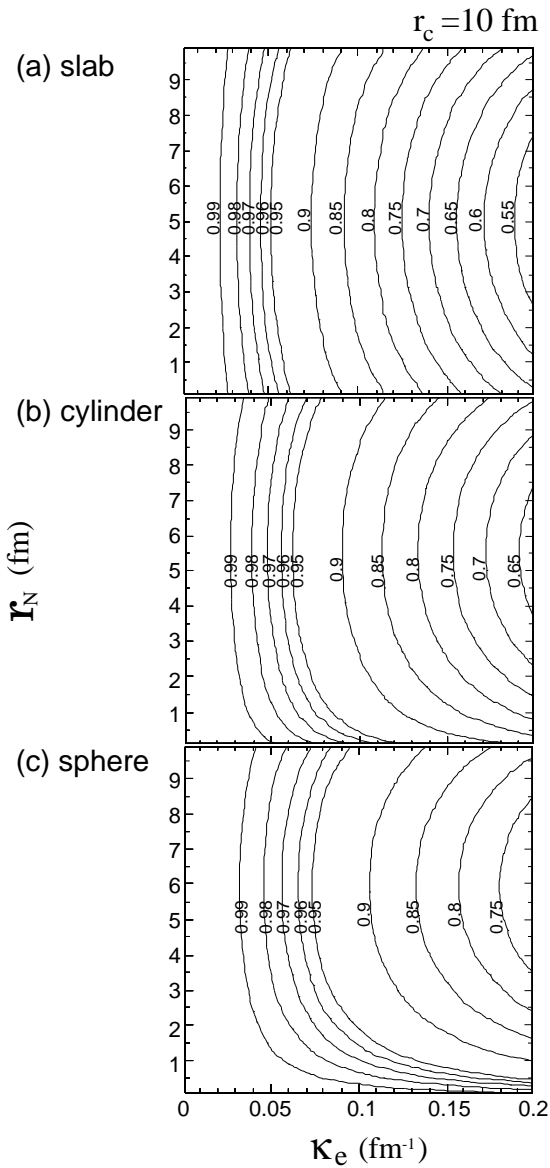}}}%
\caption{\label{ratio}
  Contour plot of the ratio $f_{d}^{\rm (screen)}/f_{d}$
  on the $\kappa_{e}$ versus $r_{\rm N}$ plane for
  (a) slablike nuclei ($d=1$),
  (b) cylindrical nuclei or bubbles ($d=2$), and
  (c) spherical nuclei or bubbles ($d=3$).
  The Wigner-Seitz cell radius $r_{\rm c}$ is fixed at $r_{\rm c}=10$ fm.
  In the Coulomb limit of $\kappa_{e} \rightarrow 0$, 
  the ratio reduces to unity.}
\end{center}
\end{figure}
\begin{figure}[htbp]
\begin{center}
\rotatebox{0}{\hspace{-10mm}
\resizebox{10cm}{!}
{\includegraphics{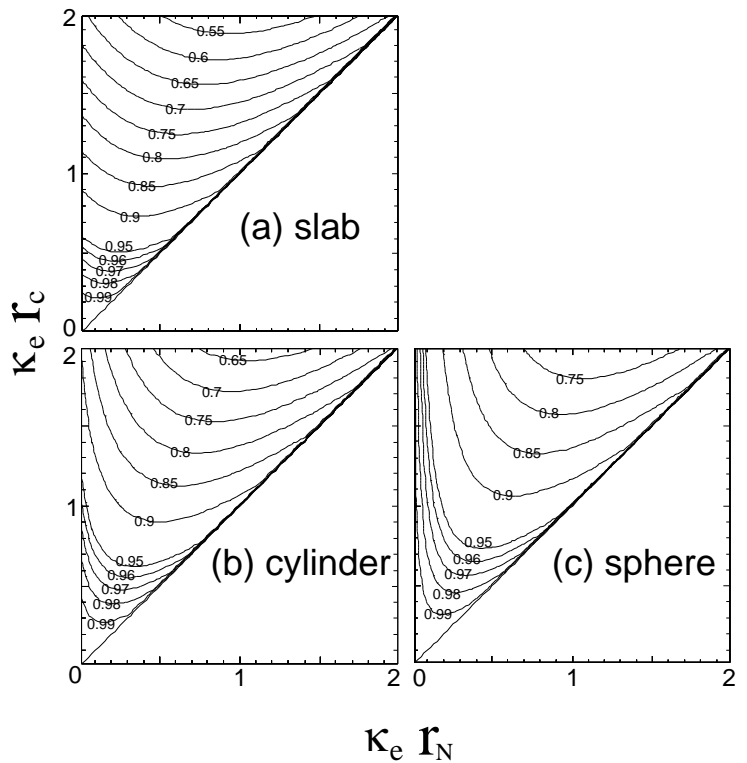}}}%
\caption{\label{ratio_kr}
  Contour plot of the ratio  $f_{d}^{\rm (screen)}/f_{d}$
  on the $\kappa_{e} r_{\rm N}$ versus $\kappa_{e} r_{\rm c}$ plane for
  (a) slablike nuclei ($d=1$),
  (b) cylindrical nuclei or bubbles ($d=2$), and
  (c) spherical nuclei or bubbles ($d=3$).}
\end{center}
\end{figure}

      In Figs.\ \ref{ratio} and \ref{ratio_kr}, we plot the ratio
$f_{d}^{\rm (screen)}(\kappa_{e}r_{\rm c}, \kappa_{e}r_{\rm N})/f_{d}(u)$,
where $f_{d}(u)$ is the non-screening limit given by
\begin{equation}
  f_d(u)= \frac{1}{d+2}
  \left[\frac{2}{d-2} \left( 1-\frac{d u^{1-2/d}}{2} \right) +u \right]\ .
\end{equation}
This expression can be obtained by taking $\kappa_{e}\to0$ in Eqs.\ 
(\ref{f1})--(\ref{f3}).  We find from Fig.\ \ref{ratio} that as $\kappa_{e}$ 
increases (i.e., the electron density $n_{e}^{(0)}$ increases), the ratio 
$f_{d}^{\rm (screen)}/f_{d}$ decreases most rapidly for the slablike 
configuration while most slowly for the spherical configuration.  This 
configuration dependence holds also for increase in $r_{\rm N}$ and 
$r_{\rm c}$, as can be seen in Fig.\ \ref{ratio_kr}.  We may thus conclude 
that for fixed values of $r_{\rm N},\ r_{\rm c}$, and $n_{e}^{(0)}$, the 
screening-induced reduction in the electrostatic energy is larger for the 
structure of lower dimensionality.

\begin{figure}[htbp]
\begin{center}
\rotatebox{0}{
\resizebox{8.2cm}{!}
{\includegraphics{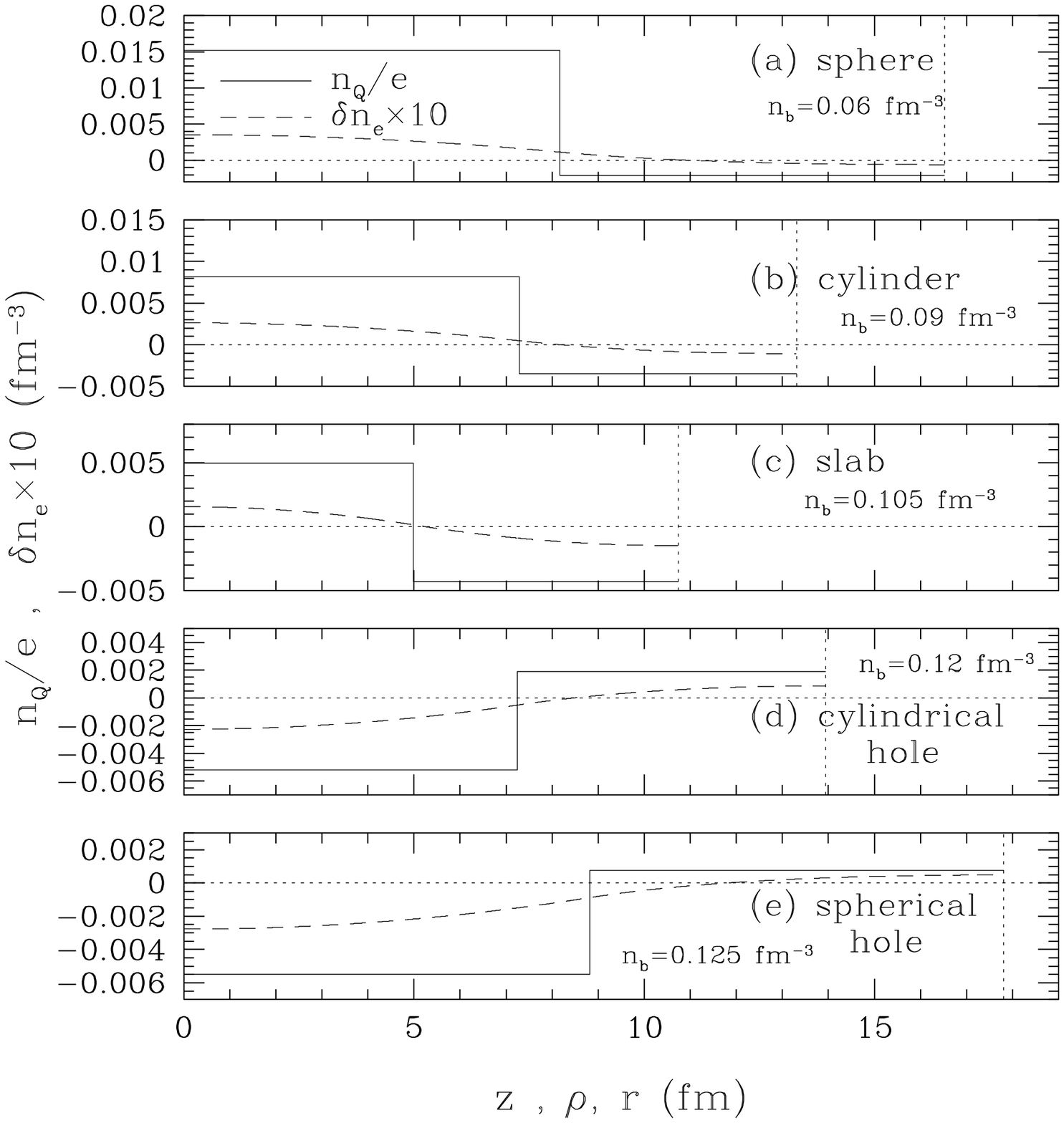}}}%
\caption{\label{distribution nsm}
  Charge distribution in a Wigner-Seitz cell in neutron star matter for the 
phases with (a) spherical nuclei, (b) cylindrical nuclei, (c) slablike nuclei,
(d) cylindrical bubbles, and (e) spherical bubbles, calculated at typical 
baryon densities for $C_2=1.0$.  The solid lines denote the unperturbed local 
charge density $n_{Q}({\bf r})$, divided by $e$, and the dashed lines denote 
the screening-induced deviation $\delta n_{e}({\bf r})$ of the electron number 
density, multiplied by ten.  The vertical dotted lines are the cell boundaries.
}
\end{center}
\end{figure}

\begin{figure}[htbp]
\begin{center}
\rotatebox{0}{
\resizebox{8.2cm}{!}
{\includegraphics{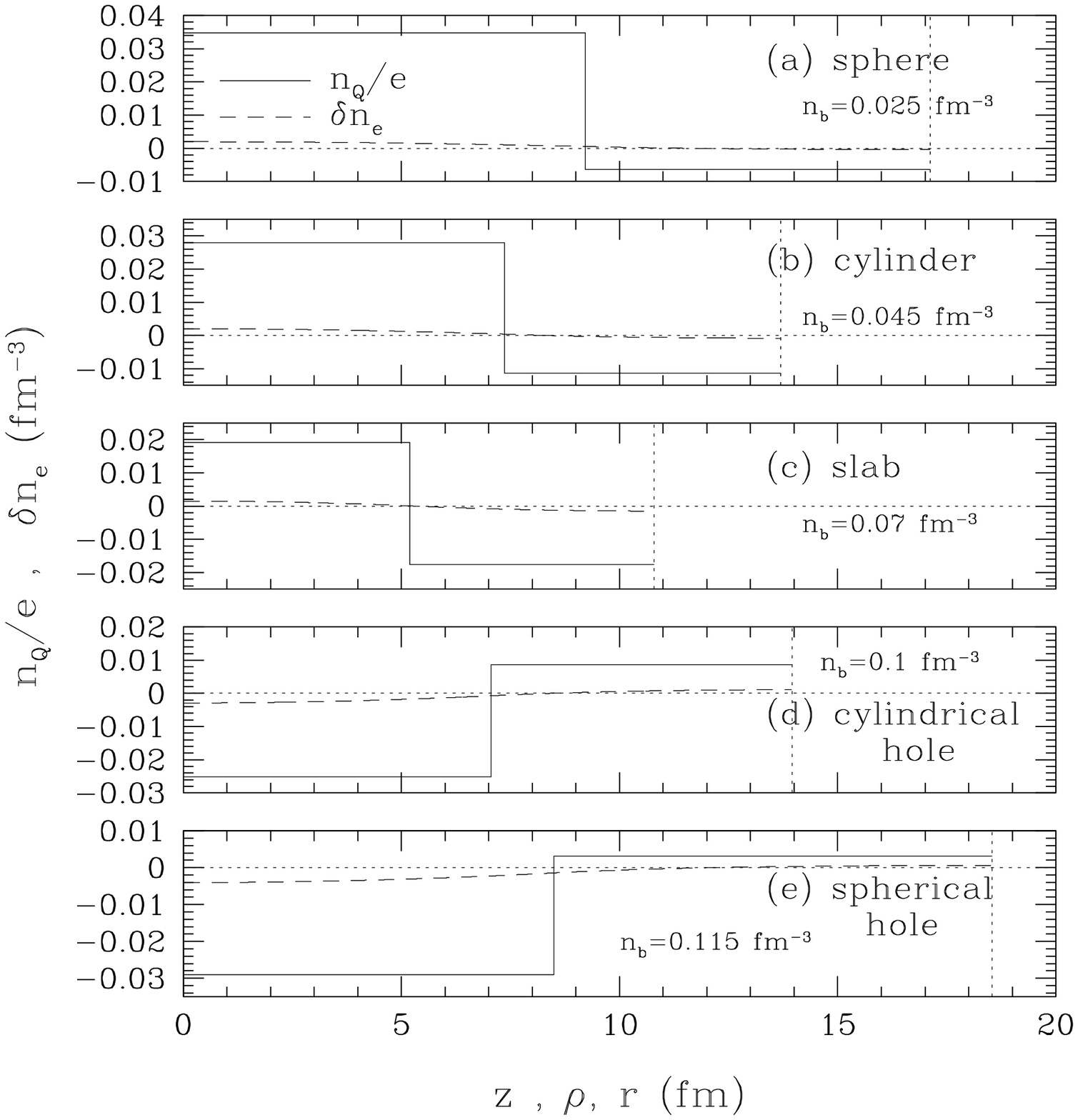}}}%
\caption{\label{distribution snm}
  Charge distribution in a Wigner-Seitz cell in supernova matter for the 
phases with (a) spherical nuclei, (b) cylindrical nuclei, (c) slablike nuclei,
(d) cylindrical bubbles, and (e) spherical bubbles, calculated at typical 
baryon densities for $C_2=1.0$ and $Y_{\rm L}=0.3$.  The solid lines denote 
the unperturbed local charge density $n_{Q}({\bf r})$, divided by $e$, and the
dashed lines denote the screening-induced deviation $\delta n_{e}({\bf r})$ of
the electron number density.  The vertical dotted lines are the cell 
boundaries.
}
\end{center}
\end{figure}

     In Figs.\ \ref{distribution nsm} and \ref{distribution snm}, we 
illustrate the charge distributions for the five crystalline phases in neutron
star matter and supernova matter, respectively.  We can see from these figures
that in both cases the magnitude of the electron density deviation 
$|\delta n_{e}|$ induced by the screening is much smaller than that of the 
unperturbed charge number density $|n_{Q}|/e$.  This ensures the validity of 
the linearized Thomas-Fermi approximation adopted here.

\section{\label{result} Screening Corrections}

\begin{figure}[htbp]
\begin{center}
\rotatebox{0}{
\resizebox{8.2cm}{!}
{\includegraphics{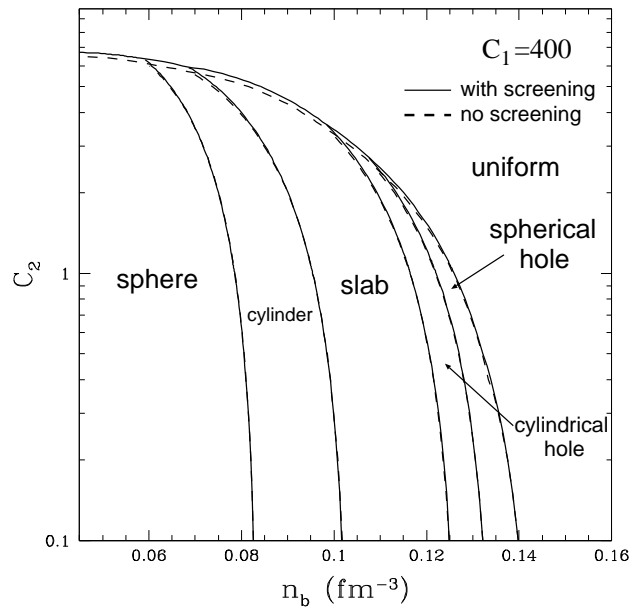}}}%
\caption{\label{phasediagram nsm}
  Zero-temperature phase diagram of neutron star matter
  on the $n_{\rm b}$ versus $C_{2}$ plane.
  The solid lines are the phase boundaries obtained for the case allowing
  for the electron screening.  The dashed lines are for the case ignoring
  the electron screening, which are taken from the lower left panel
  in Fig.\ 3 in Ref.\ \cite{gentaro1}.}
\end{center}
\end{figure}

\begin{figure}[htbp]
\begin{center}
\rotatebox{0}{
\resizebox{8.2cm}{!}
{\includegraphics{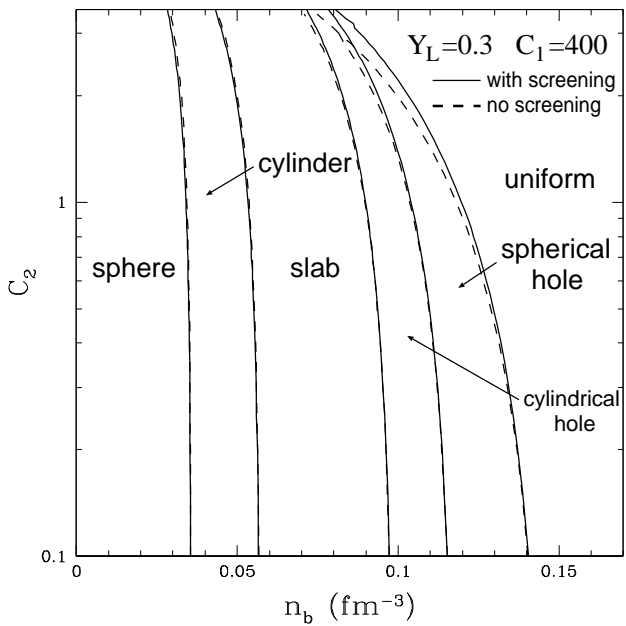}}}%
\caption{\label{phasediagram snm}
  Zero-temperature phase diagram of supernova matter
  on the $n_{\rm b}$ versus $C_{2}$ plane.
  The solid lines are the phase boundaries obtained for the case allowing
  for the electron screening.  The dashed lines are for the case ignoring
  the electron screening, which are taken from the lower left panel
  in Fig.\ 3 in Ref.\ \cite{gentaro2}.}
\end{center}
\end{figure}

     We now examine the influence of the electron screening on the phase 
diagrams of the ground-state neutron star matter and supernova matter at 
subnuclear densities.  In Figs.\ \ref{phasediagram nsm} and 
\ref{phasediagram snm}, we draw the phase diagrams of neutron star matter and 
supernova matter, respectively, on the $n_{\rm b}$ versus $C_{2}$ plane for 
the cases with and without screening.  As can be seen from these figures, the 
screening 
leads to slight expansion of the density region in which the ``pasta'' phases
appear, and this expansion 
is larger in supernova matter than in neutron star matter.  

\begin{figure}[htbp]\vspace{5mm}\hspace{-5mm}
\begin{center}
\rotatebox{0}{
\resizebox{8.7cm}{!}
{\includegraphics{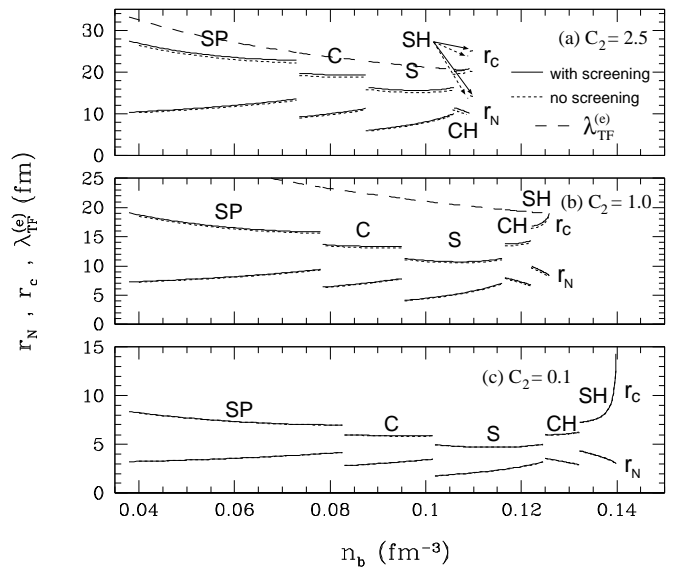}}}%
\caption{\label{size nsm}
  Size of a nucleus or bubble, $r_{\rm N}$, and of a Wigner-Seitz cell, 
  $r_{\rm c}$, in neutron star matter
  calculated for $C_{2}=0.1$, 1.0, and 2.5.
  The Thomas-Fermi screening length $\lambda_{\rm TF}^{(e)}$ is also plotted.
  The solid lines are the results for the case with screening, and the dashed 
  lines are the results for the case without screening, which are taken from
  Fig.\ 5 in Ref.\ \cite{gentaro1}.
  The symbols SP, C, S, CH, and SH stand for 
  sphere, cylinder, slab, cylindrical hole, and spherical hole, 
  respectively.}
\end{center}
\end{figure}

\begin{figure}[htbp]\vspace{5mm}\hspace{-5mm}
\begin{center}
\rotatebox{0}{
\resizebox{8.7cm}{!}
{\includegraphics{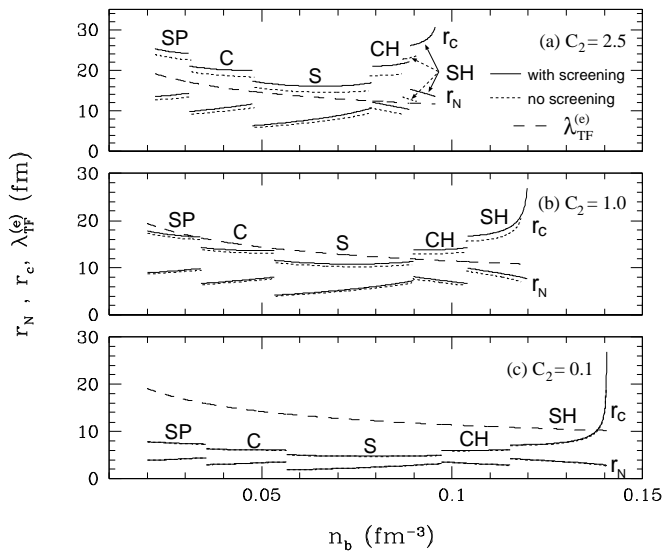}}}%
\caption{\label{size snm}
  Size of a nucleus or bubble, $r_{\rm N}$, and of a Wigner-Seitz cell, 
  $r_{\rm c}$, in supernova matter
  calculated for $Y_{\rm L}=0.3$ and $C_{2}=0.1$, 1.0, and 2.5.
  The Thomas-Fermi screening length $\lambda_{\rm TF}^{(e)}$ is also plotted.
  The solid lines are the results for the case with screening, and the dashed 
  lines are the results for the case without screening, which are taken from
  Fig.\ 5 in Ref.\ \cite{gentaro2}.
  The symbols SP, C, S, CH, and SH stand for 
  sphere, cylinder, slab, cylindrical hole, and spherical hole, 
  respectively.}
\end{center}
\end{figure}

     In order to consider why the screening is more effective in supernova 
matter than in neutron star matter, it is useful to plot the screening length 
$\lambda_{\rm TF}^{(e)}$ and the spatial scales, $r_{\rm N}$ and $r_{\rm c}$, 
as in Figs.\ \ref{size nsm} and \ref{size snm}.  By comparing these figures, 
we can see that the ratios $r_{\rm N}/\lambda_{\rm TF}^{(e)}$ and 
$r_{\rm c}/\lambda_{\rm TF}^{(e)}$, whose increase makes the electron 
screening more effective, is larger in supernova matter than in neutron star 
matter.  This is because of 
smaller $\lambda_{\rm TF}^{(e)}$
in supernova matter.  

     We also see from  Figs.\ \ref{size nsm} and \ref{size snm} that 
the electron screening acts to increase  $r_{\rm N}$ and $r_{\rm c}$ by a 
small amount, while keeping the ratio $r_{\rm N}/r_{\rm c}$, or equivalently, 
$u$, almost unchanged.
Such increase in $r_{\rm N}$ stems from the fact that as discussed in Sec.\ 
\ref{model}, the electron screening modifies the condition for size 
equilibrium from Eq.\ (\ref{usual}) to Eq.\ (\ref{sizeeq}), which can be 
rewritten as
\begin{equation}
  w_{\rm surf}=2(w_{\rm C+L,0} + \delta w_{\rm C+L}
    + \delta E_e)\ . \label{sizeeq mod}
\end{equation}
Here, $w_{\rm C+L,0}=2\pi(nxe)^2r_{\rm N}^2 u f_d(u)$ is the total 
electrostatic energy density in the non-screening limit, and 
$\delta w_{\rm C+L}$ and $\delta E_e$ are the corrections 
to Eq.\ (\ref{usual})
due to the 
screening-induced changes in the electrostatic energy and the electron energy,
respectively, which are proportional to $\kappa_e^2$ in leading order.
Up to $O(\kappa_e^2)$, the sum $\delta w_{\rm C+L}$ and $\delta E_e$ is 
negative since $\delta w_{\rm C+L}$, which is negative, is more important 
than $\delta E_e$, which is positive.  
Consequently, the screening makes the equilibrium value of $r_{\rm N}$ larger 
than that in the non-screening limit, given by
\begin{equation}
 r_{\rm N}^{(0)}=\left[\frac{d E_{\rm surf}}{4\pi(nxe)^2 f_d}\right]^{1/3}\ . 
\label{eqsize}
\end{equation}
On the other hand, the negligible screening effect on $u$ suggests that the 
pressure corrections due to the screening through the Coulomb pressure and 
the electron pressure [see the last two terms in the right side of Eq.\ 
(\ref{pn})] are negligibly small.

     We proceed to see how the phase structure changes with the strength of 
the surface tension, $C_{2}$, and the baryon density $n_{\rm b}$.  We find 
from Figs.\ \ref{phasediagram nsm} and \ref{phasediagram snm} that as $C_{2}$ 
decreases, the phase boundaries in the case with screening approach those in 
the case without screening.  This is consistent with the fact that for weaker 
surface tension, the equilibrium size of the spatial structure becomes smaller,
leading to smaller $r_{\rm N}/\lambda_{\rm TF}^{(e)}$ and 
$r_{\rm c}/\lambda_{\rm TF}^{(e)}$.  As $n_{\rm b}$ increases with $C_{2}$ 
fixed, on the other hand, the screening induced change in the phase boundaries
becomes more appreciable; the increase in the transition density between the 
cylindrical hole to the spherical hole phase is larger than that between
the slab to the cylindrical hole phase while being smaller than that between 
the spherical hole to the uniform phase.  This is partly because 
$\lambda_{\rm TF}^{(e)}$ decreases with increasing density and partly because 
at fixed $n_{\rm b}$, the higher dimensionality has the larger equilibrium 
values of $r_{\rm N}$ and $r_{\rm c}$.  This dimensionality dependence, which 
was also obtained in earlier investigations based on various nuclear models 
(see, e.g., Refs.\ \cite{lorenz,oyamatsu,gentaro1,gentaro2}), stems from the 
fact that generally the equilibrium values of the surface energy density 
$w_{\rm surf}$ and surface tension $E_{\rm surf}$ are almost degenerate among 
the five crystalline phases at fixed $n_{\rm b}$ and thus the equilibrium 
value of $r_{\rm N}$ behaves roughly as $r_{\rm N}\propto d$ [see Eq.\ 
(\ref{esurf})].

     In order to examine the influence of 
the screening on the phase 
boundaries in further detail, we list in Table \ref{density} the transition 
densities calculated for neutron star matter and supernova matter at a typical
value of $C_{2}=1.0$.  We thus find that the ``pasta'' phases, as a whole, 
are slightly enlarged by the electron screening in such a way that the phase 
boundaries associated with the bubble phases move into a higher density, 
whereas the others move into a lower density.  Such movement stems from the 
dimensionality dependence of the quantity, $f_{d}^{\rm (screen)}/f_{d}$, 
characterizing 
the screening correction to the electrostatic energy: At fixed $r_{\rm N}$, 
$r_{\rm c}$, and $n_{e}^{(0)}$, as was found from Figs.\ \ref{ratio} and 
\ref{ratio_kr}, $f_{d}^{\rm (screen)}/f_{d}$ is smaller and hence the 
screening is more efficient for lower dimensionality.  
We remark in passing that the screening correction to the electrostatic 
energy dominates over the screening correction to the electron energy.

\begin{table}
  \caption{\label{density} Transition densities (in fm$^{-3}$)
    in neutron star matter (NSM) and supernova matter (SNM), calculated 
    at $C_{2}=1.0$.  The lepton fraction $Y_{\rm L}$ is set at 0.3
    for supernova matter.  The symbols $\bigcirc$ and $\times$ denote
    the cases with and without screening, respectively.
    The symbols SP, C, S, CH, SH, and U
    stand for sphere, cylinder, slab, cylindrical hole, spherical hole, and
    uniform matter, respectively.}
\begin{ruledtabular}
\begin{tabular}{ccccccc}
  &{\rm screening}&
  {\rm SP}$\leftrightarrow${\rm C} & {\rm C}$\leftrightarrow${\rm S} &
  {\rm S}$\leftrightarrow${\rm CH} & {\rm CH}$\leftrightarrow${\rm SH} &
  {\rm SH}$\leftrightarrow${\rm U}\\
  {\rm NSM} & $\times$ &
  0.07854 & 0.09537 & 0.11636 & 0.12191 & 0.12571\\
  {\rm NSM} & $\bigcirc$ &
  0.07849 & 0.09535 & 0.11648 & 0.12206 & 0.12592\\
  {\rm SNM} & $\times$ &
  0.03433 & 0.05364 & 0.08965 & 0.10364 & 0.11812\\
  {\rm SNM} & $\bigcirc$ &
  0.03407 & 0.05334 & 0.08990 & 0.10410 & 0.11974\\
\end{tabular}
\end{ruledtabular}
\end{table}

     The decrease in the transition density between the phases with spherical 
and with cylindrical nuclei can be understood from the condition for fission 
instability of spherical nuclei in a Coulomb lattice.  Up to $O(u^{1/3})$, 
this condition reads \cite{iida}
\begin{equation}
  w_{\rm C} \geq 2w_{\rm surf}\ ,\label{fission}
\end{equation}
where $w_{\rm C}=4\pi(nxe)^2 u r_{\rm N}^2/5$ is the Coulomb self energy 
of the nucleus, divided by the cell volume.  Condition (\ref{fission}) 
provides a critical size, 
\begin{equation}
r_{\rm N}^{\rm crit}=\left[\frac{30 E_{\rm surf}}{4\pi(nxe)^2}\right]^{1/3}\ .
 \label{critsize}
\end{equation}
In the absence of screening, the equilibrium nuclear size $r_{\rm N}^{(0)}$, 
given by Eq.\ (\ref{eqsize}) in which $f_{d=3}$ is retained up to 
$O(u^{1/3})$, reaches this critical size when $u=1/8$.  In the presence of 
screening, as in Figs.\ \ref{size nsm} and \ref{size snm}, the equilibrium 
value of $r_{\rm N}$ is larger than $r_{\rm N}^{(0)}$ at given $n_{\rm b}$.  
Since the equilibrium value of $r_{\rm N}$ increases with $n_{\rm b}$ for 
spherical nuclei, the density region in which the spherical nuclei are 
unstable against fission extends to a lower density.

     The increase in the density at which the system turns into uniform matter
is the most salient feature induced by the screening; as can be seen from 
Table \ref{density}, the density increase
for typical values of the surface tension ($C_{2}\simeq 1$) amounts to
$\sim 2\times 10^{-4}$ fm$^{-3}$ for neutron star matter and $\sim 0.0015$
fm$^{-3}$ for supernova matter.  This feature can be understood from the 
condition for proton clustering instability in $\beta$-equilibrated uniform 
nuclear matter, which was originally obtained by BBP \cite{bbp} for neutron 
star matter.  
(For supernova matter, one can follow the same argument as that 
for neutron star matter since the presence of trapped neutrinos makes no 
difference in the argument.)
They expanded the energy density 
functional $E[n_{i}({\bf r})]$ $(i=n,\ p,\ e)$ in the extended Thomas-Fermi 
model with respect to small proton density fluctuations $\delta n_{p}({\bf r})$
around the homogeneous state.  The variation of the total energy caused by the 
inhomogeneity yields
\begin{equation}
  E-E_{0}=\frac{1}{2} \int \frac{d^{3}{\bf q}}{(2\pi)^{3}}
     v(q) |\delta n_{p}({\bf q})|^{2}\ ,
\end{equation}
where $E_{0}$ is the energy of the unperturbed state, $\delta n_{p}({\bf q})$ 
is the Fourier transform of $\delta n_{p}({\bf r})$, and $v(q)$ is the 
effective potential between protons given by
\begin{equation}
  v(q) = v_{0} + \beta q^{2}
  + \frac{4\pi e^{2}}{q^{2}+\kappa_e^{2}}\ .
\label{veff}
\end{equation}
The first term in the right side of Eq.\ (\ref{veff}) is the bulk contribution,
and the second is the gradient contribution (for detailed expressions for 
$v_{0}$ and $\beta$, see Ref.\ \cite{bbp}).  The potential $v(q)$ takes a 
minimum value $v_{\rm min}$ at $q=Q$, where 
\begin{eqnarray}
  Q^{2} &=& \left( \frac{4\pi e^{2}}{\beta} \right)^{1/2}
    - {\kappa_e}^{2}\ ,\\
  v_{\rm min} &=& v_{0} + 2(4\pi e^{2}\beta)^{1/2}
    - \beta {\kappa_e}^{2}\ .
  \label{vmin}
\end{eqnarray}
The condition for the proton clustering instability, $v_{\rm min} \leq 0$, is
satisfied at densities below a critical density $n_{\rm inst}$.  This is 
because the bulk contribution $v_{0}$, which is dominant in $v_{\rm min}$, is 
an increasing function of $n_{\rm b}$ at densities around $v_{0},\ v_{\rm min} 
\sim 0$ (see, e.g., Fig.\ 2 in Ref.\ \cite{iida}).  In the presence of the 
electron screening the critical density $n_{\rm inst}$ becomes higher since
the Coulomb interaction between charge inhomogeneities, which tends to 
suppress the proton clustering, is weakened by the electron screening through
the term $-\beta{\kappa_e}^{2}(<0)$ in Eq.\ (\ref{vmin}).

     In summary we have examined the electron screening in the inhomogeneous 
phases of nuclear matter and clarified its influence on the zero-temperature 
phase diagram of neutron star matter and supernova matter at subnuclear 
densities.  We have found that the density region occupied by the ``pasta'' 
phases is slightly expanded by the electron screening in such a way that the 
three phase boundaries associated with the bubble phases move into a higher 
density, whereas the other two move into a lower density.  This expansion 
stems from the model independent features that the electron screening makes
spherical nuclei more subject to fission by enlarging the equilibrium size 
through modifications of the size equilibrium condition as in Eq.\ 
(\ref{sizeeq mod}) and that the role played by the Coulomb interaction between 
small charge inhomogeneities in uniform nuclear matter in suppressing the 
proton clustering instability is weakened by the electron screening.

     Finally, we consider on what physical conditions a mixed state composed 
of two phases having different charge density appear as in the nuclear 
liquid-gas mixed state studied here.  As mentioned in Ref.\ \cite{heiselberg} 
in the context of a quark-hadron mixed state, a mixed state can be stable only
when the charge screening and surface tension are sufficiently weak.  In the
present nuclear case, the charge screening is weak and mainly reduces the
electrostatic energy, leading to a larger density range of the mixed state
for typical values of the surface tension.  In the case of the quark-hadron 
mixed state, the charge screening can be strong enough to raise the bulk 
energy of the system and hence prefer the phase separated state over the mixed
state for realistic values of the surface tension, as discussed in Ref.\ 
\cite{tatsumi}.

\begin{acknowledgments}

The authors are grateful to T. Tatsumi for insightful discussions and comments.
G.W. is also grateful to K. Sato, K. Yasuoka, and T. Ebisuzaki for continuous 
encouragement.  This work was supported in part by RIKEN Junior Research 
Associates Grant No.\ J130026, by a Grant-in-Aid for Scientific Research 
provided by the Ministry of Education, Culture, Sports, Science, and 
Technology of Japan through Grant No.\ 14-7939, and by RIKEN Special 
Postdoctoral Researchers Grant No.\ 011-52040.

\end{acknowledgments}

%\bibliography{qmd1}

\end{document}